# Hypersymmetry field rotation angle

György Darvas[*]

Symmetrion[#] and Faculty of Sciences, Eötvös Loránd University, Budapest, Hungary

http://orcid.org/0000-0002-1652-4138

darvasg@caesar.elte.hu

**Abstract**

The paper determines a limit energy under which hypersymmetry (HySy) is broken. According to gauge theories, interaction mediating spin-0 bosons must be massless. The theory of HySy predicted massive intermediate bosons. Hypersymmetry field rotation, described in this paper, justifies the mass of the HySy mediating boson. The mass of intermediate bosons must arise from dynamical spontaneous breaking of the group of HySy. HySy rotation is performed in the velocity-dependent **D** field. The derived rotation of the field is defined by the spontaneous symmetry breaking and precession of the velocity *v* around its third projection in the **D** field (that produced the mass of the field's bosons). The latter represents the real- and effective velocities of a boson-emitting particle in the direction towards a target particle. The mass of the discussed (fictitious) Goldstone bosons can be removed by the unitarity gauge condition through Higgs (BEH) mechanism. According to the simultaneous presence of a Standard Model (SM) interaction's symmetry group and the (beyond SM) HySy group, their bosons should be transformed together. Spontaneous breakdown of HySy may allow performing a transformation that does not influence the SM physical state of the investigated system. The paper describes a field transformation that eliminates the mass of the intermediate bosons, rotates the SM- and HySy bosons' masses together while leaving the SM bosons intact. The result is an angle that characterises the HySy by a precession mechanism of the velocity that generates the field. In contrast to the known SM intermediate bosons, the HySy intermediate bosons have no fixed mass. The mass of the HySy intermediate bosons (that appear as quanta of a velocity-dependent gauge field **D**) depends on the relative velocity of the particles whose interaction they mediate. So, the derived precession angle is a function of that velocity.

[*] Retired from the Faculty of Sciences, Eötvös Loránd University, and the IRO of the Hungarian Academy of Sciences.

[#] This paper was prepared in the framework of a long-term bilateral cooperation project between the *Hungarian* and *Russian Academies of Sciences* under the item 5 in the complex working plan entitled *Non-linear models and symmetry analysis in* biomechanics, bioinformatics, and *the theory of self-organizing systems*.

The work was sponsored also by the grant 002/2018 of the Symmetrology Foundation.

# 1 Introduction

Theories are generally considered relativistic if they meet the condition to be invariant under Lorentz transformation. E.g., General Theory of Relativity (GTR) and Quantum Electrodynamics (QED) both meet this condition. However, derivations of all the Einstein equations and their solutions, as well as the Dirac equation (and related other discussions of QED) are performed with assuming preliminary approximations to "not too high velocities". In fact, invariance under the Lorentz transformation is a necessary condition for a theory to be relativistic [1], but that condition is not (always) sufficient [2], [3]. High energy experiments reach speeds very near to that of the light. In the interpretation of their results, one can disregard the effects of those high velocities no more. The absence of such precise high-velocity considerations led to anomalies that – among others – formulated the demand for extending the Standard Model (SM) in the nineties. In contrast to other models proposed for the extension of the SM, hypersymmetry (HySy) [1] offered an alternative by the application of a strongly relativistic model [6], [7], [8], [9], [10], [11], [12], [13], [3]. This model includes (among others) a velocity-dependent field (**D**), which proved to be a gauge field, and should be added to the SM fields. The assumption of **D** is no more surprising than the introduction of the Higgs (BEH) field was in the mid-1960ies. Accepting this assumption, prediction of intermediate bosons of the **D** field could already have indicated less surprise.

According to gauge theories, interaction mediating bosons must be massless. The theory of HySy predicted spin-0 but massive intermediate bosons ($\delta$). The mass of intermediate bosons in spin-0 fields must arise from dynamical symmetry breaking [14], [15], [16], [17]. The mass of $\delta$ should arise from spontaneous breaking of the group of HySy. The group of HySy has two free parameters. Its spontaneous breakdown may eliminate one of them: it allows to perform a transformation that does not influence the physical state of the investigated system. The other free parameter can be discussed in terms of the BEH (Brout-Englert-Higgs) mechanism [20], [21]. The intermediate bosons of the SM belong to one of the three Goldstone boson types defined by Weinberg[2]. The simultaneously appearing HySy $\delta$ boson belongs to the fictitious Goldstone bosons, whose mass is removed by the Higgs mechanism. The mass of the fictitious Goldstone bosons is eliminated by the unitarity gauge condition. (HySy meets that condition). According to the simultaneous presence of a SM interaction's symmetry group and the HySy, their bosons should be rotated simultaneously ($G_{SM} \otimes G_D$)[3].

---

[1] The term *hypersymmetry* was used in physics in the mid-nineties last time, although with quite a different meaning. Its appearance was related to the childhood of supersymmetry in the sense of meaning an extension, while in other publications, the idea of hypersymmetry was used as generalisation of supersymmetry (first of all by R. Kerner *et al.* [4], [5]). That meaning was shortly abandoned and the proposed algebra was not used for long. We have applied this term in a new meaning for an alternative, competing model of the supersymmetry.

[2] Originally, it was assumed that the spontaneous symmetry breakdown responsible for the intermediate vector-boson masses was due only to the vacuum expectation values of a set of spin-0 fields. Later this approach became more sophisticated, and it was assumed that the considered symmetry breaking was of a purely dynamical nature. Weinberg [18] distinguished three types of Goldstone bosons (fictitious, true and pseudo-), and, accordingly, three dynamical symmetry breaking mechanisms. We remark that Weinberg's classification allowed the existence of other (at that time unknown) gauge fields and intermediate bosons, which encouraged the elaboration of the HySy field rotation mechanism discussed here. In respect of the latter, cf., also the remarks Sec. IV. (3)(B) in [16]. Other aspects of the problem are offered at [19].

[3] Comparison of the rotation of the $\delta$ in combination with a respective SM boson, and the mixing of the also massive neutral weak vector boson with $\gamma$ needs further investigation. Note, the latter are simultaneously rotated (by $\theta_W$) in field **B**, while the former are rotated in field **D** (characterised indirectly by $\theta_D$) as we will see below.

In the course of interactions between two fermions, the bosons of the **D** field are expected to be exchanged (anti/)parallel with the exchange of a SM boson [7], since the **D** field always appears as an extension to a SM interaction field. The HySy theory must try to avoid affecting the respective SM bosons. For this reason, when the HySy theory assumes that the dions obtain their mass by a transformation of the non-SM **D** field via a BEH-mechanism, it should guarantee that the respective SM boson be left intact. The latter condition demands the existence of a transformation matrix that includes a particular transformation angle (like the fermion flavour mixing $\theta_C$ Cabibbo angle [22]). This rotation – that can be characterised by an inclination angle $\theta_D$ – can guarantee that an expected 0 mass (Goldstone) boson would transform into the predicted mass HySy $\delta$ boson (or back), while the (anti/)parallel SM boson does not change its properties, similar to the mechanism of the electro-weak theory's weak (Weinberg) mixing angle (cf., footnote 2). The existence of such a HySy field rotation mechanism is discussed below.

## 2 Mass of the $\delta$ boson

The interaction between two isotopes of any field-charge [6], [7] is mediated by a massive non-SM gauge boson. According to the isotopic field-charge (IFC) theory, the mass of this boson (called dion, $\delta$) is the difference between the boosted (dressed) and the rest (bare or invariant) masses. The mass of this gauge boson is independent of the type (gravitational, electromagnetic, weak, strong) of the interaction: $m(\delta) = m_T - m_V = (\kappa - 1)m_V$ where $\kappa = 1/\sqrt{1-(v/c)^2}$, $m_T$ is the Lorentz boosted mass and $m_V$ denotes the mass that appears in the potential (scalar) part of the Hamiltonian, which is equal to the rest mass of the concerned field-charge.

The presence of a massive mediating boson assumes a spontaneous symmetry breaking. Spontaneous symmetry breaking rotates the massless Goldstone boson plane, producing, as a result, the massive $\delta$ boson and the respective SM mass bosons (here denoted by $\xi$). In the opposite direction, the same rotation transforms (in its gauge field) the massive HySy $\delta$ boson's and the respective SM $\xi$ boson's plane into a massless Goldstone boson ($\delta'$) while leaving the SM $\xi$ boson intact [cf., Eqs. (11)+]. In the instance of the isotopic field-charge field (marked **D**), the quanta of this field ($\delta$) are associated with the conservation of the isotopic field-charge spin (IFCS or $\Delta$) introduced in physical terms in [7]. The (inverse) transformation that eliminates "unwanted" masses produced by the spontaneous symmetry breaking is expected to depend on the velocity of the interacting isotopic field-charges relative to each other and assumes the presence of a velocity-dependent (kinetic) *field* (**D**) instead of simple configuration *space*.

This paper defines the transformation of the fields, whose quanta are the $\delta$ bosons. Note that the $\delta$ bosons never appear alone. They act simultaneously (parallel or antiparallel) with one of the Standard Model (SM) bosons. Therefore, one requires the transformation of the **D** field together with one of the SM fields (denoted here by $\mathbf{X}_{SM}$). Note also that the derivation of the field equations of the interactions and their solutions included approximations (cf. Sec. 1 above). Although all field theories required invariance under the Lorentz transformation, they included restrictions to "not too high" velocities; meaning that those approximations cannot be applied at the high kinetic energies (and the respective high velocities) for the interpretation of data collected in experiments producing large accelerations. (The velocity limit is discussed in Section 4 of this paper.)

## 3 Transformation in a coupled SM field and the D field – The origin of the mass of *δ*

Conservation [23] of the *Δ* quantity first requires *invariance under hypersymmetry* (HySy, cf., [3]). At the same time, the interaction between two particles requires *invariance under the Lorentz transformation*. As it was shown several times (among others, in [3]), invariance of physical theories under the Lorentz transformation is a necessary condition, but it is not always sufficient. In certain instances (cf. [3]), the transformation needs to be complemented with others.

A general form of the Lorentz transformation's matrix can be written as follows:

$$\Lambda = \begin{bmatrix} 1+(\kappa-1)\frac{v_1^2}{v^2} & (\kappa-1)\frac{v_1 v_2}{v^2} & (\kappa-1)\frac{v_1 v_3}{v^2} & i\kappa\frac{v_1}{v}\frac{v}{c} \\ (\kappa-1)\frac{v_2 v_1}{v^2} & 1+(\kappa-1)\frac{v_2^2}{v^2} & (\kappa-1)\frac{v_2 v_3}{v^2} & i\kappa\frac{v_2}{v}\frac{v}{c} \\ (\kappa-1)\frac{v_3 v_1}{v^2} & (\kappa-1)\frac{v_3 v_2}{v^2} & 1+(\kappa-1)\frac{v_3^2}{v^2} & i\kappa\frac{v_3}{v}\frac{v}{c} \\ -i\kappa\frac{v_1}{v}\frac{v}{c} & -i\kappa\frac{v_2}{v}\frac{v}{c} & -i\kappa\frac{v_3}{v}\frac{v}{c} & \kappa \end{bmatrix} \quad (1)$$

where *v* is the velocity of the interacting particles relative to each other; $v_i$ are the components of *v*; and $v_1^2 + v_2^2 + v_3^2 = v^2$. This formula holds when the origin of the reference frame is fixed to one of the interacting field-charges and restricted to the situation when the velocity vector arrows from one of the field-charges towards the other (at least, while the velocity is not too high, as we will demonstrate it following Eq. (18)). At this stage, we do not require any prescription for the direction of the co-ordinate axes. Let's interpret the velocities that define the Lorentz transformation in the configuration space, transformed into the above-mentioned velocity-dependent field.

Let's introduce the following notations:

$$\frac{v}{c} = \sin\vartheta; \quad \kappa = \frac{1}{\cos\vartheta}; \quad (\kappa - 1) = \frac{1-\cos\vartheta}{\cos\vartheta}; \quad \text{and } u_i = \frac{v_i}{v} \ (i = 1, 2, 3).$$

The $u_i$ are unitary length ($u_1^2 + u_2^2 + u_3^2 = 1$) vector components, pointing in the direction of the axes of the coordinate system. So, the Lorentz transformation can be rewritten in the following forms:

$$\Lambda = \begin{bmatrix} 1+\dfrac{1-\cos\vartheta}{\cos\vartheta}u_1^2 & \dfrac{1-\cos\vartheta}{\cos\vartheta}u_1u_2 & \dfrac{1-\cos\vartheta}{\cos\vartheta}u_1u_3 & iu_1\mathrm{tg}\,\vartheta \\ \dfrac{1-\cos\vartheta}{\cos\vartheta}u_2u_1 & 1+\dfrac{1-\cos\vartheta}{\cos\vartheta}u_2^2 & \dfrac{1-\cos\vartheta}{\cos\vartheta}u_2u_3 & iu_2\mathrm{tg}\,\vartheta \\ \dfrac{1-\cos\vartheta}{\cos\vartheta}u_3u_1 & \dfrac{1-\cos\vartheta}{\cos\vartheta}u_3u_2 & 1+\dfrac{1-\cos\vartheta}{\cos\vartheta}u_3^2 & iu_3\mathrm{tg}\,\vartheta \\ -iu_1\mathrm{tg}\,\vartheta & -iu_2\mathrm{tg}\,\vartheta & -iu_3\mathrm{tg}\,\vartheta & \dfrac{1}{\cos\vartheta} \end{bmatrix} \qquad (2)$$

$$\Lambda = \kappa \begin{bmatrix} \cos\vartheta+(1-\cos\vartheta)u_1^2 & (1-\cos\vartheta)u_1u_2 & (1-\cos\vartheta)u_1u_3 & iu_1\sin\vartheta \\ (1-\cos\vartheta)u_2u_1 & \cos\vartheta+(1-\cos\vartheta)u_2^2 & (1-\cos\vartheta)u_2u_3 & iu_2\sin\vartheta \\ (1-\cos\vartheta)u_3u_1 & (1-\cos\vartheta)u_3u_2 & \cos\vartheta+(1-\cos\vartheta)u_3^2 & iu_3\sin\vartheta \\ -iu_1\sin\vartheta & -iu_2\sin\vartheta & -iu_3\sin\vartheta & 1 \end{bmatrix} \qquad (3)$$

Let's take a general $\vartheta$ angle rotation matrix in a 3D space stretched by unit axis vectors $u_i$:

$$R = \begin{bmatrix} \cos\vartheta+(1-\cos\vartheta)u_1^2 & (1-\cos\vartheta)u_1u_2 - u_3\sin\vartheta & (1-\cos\vartheta)u_1u_3 + u_2\sin\vartheta & 0 \\ (1-\cos\vartheta)u_2u_1 + u_3\sin\vartheta & \cos\vartheta+(1-\cos\vartheta)u_2^2 & (1-\cos\vartheta)u_2u_3 - u_1\sin\vartheta & 0 \\ (1-\cos\vartheta)u_3u_1 - u_2\sin\vartheta & (1-\cos\vartheta)u_3u_2 + u_1\sin\vartheta & \cos\vartheta+(1-\cos\vartheta)u_3^2 & 0 \\ 0 & 0 & 0 & 1 \end{bmatrix} \qquad (4)$$

However, one can interpret the $R$ transformation of the velocity vector components as also projected in the velocity-dependent field. Note, while $\vartheta$ is a symbolic notation in the Lorentz transformation, it denotes a real rotation angle in $R$.[4] Let's compare this transformation matrix $R$ with the formula for the Lorentz transformation. In this order, let's decompose R to the following two matrices:

$$R = \begin{bmatrix} \cos\vartheta+(1-\cos\vartheta)u_1^2 & (1-\cos\vartheta)u_1u_2 & (1-\cos\vartheta)u_1u_3 & 0 \\ (1-\cos\vartheta)u_2u_1 & \cos\vartheta+(1-\cos\vartheta)u_2^2 & (1-\cos\vartheta)u_2u_3 & 0 \\ (1-\cos\vartheta)u_3u_1 & (1-\cos\vartheta)u_3u_2 & \cos\vartheta+(1-\cos\vartheta)u_3^2 & 0 \\ 0 & 0 & 0 & 1 \end{bmatrix} + \\ + \begin{bmatrix} 0 & -u_3\sin\vartheta & u_2\sin\vartheta & 0 \\ u_3\sin\vartheta & 0 & -u_1\sin\vartheta & 0 \\ -u_2\sin\vartheta & u_1\sin\vartheta & 0 & 0 \\ 0 & 0 & 0 & 0 \end{bmatrix} \qquad (5)$$

and in a bit extended form:

---

[4] The method to be applied shows certain partial similarity to the derivation of Wigner-Thomas rotation, which applies that a boost (here by $v$) and a rotation are equivalent with the combination of two coupled boosts. In the inverse, they correspond to transformations whose combination produces a (Thomas-) precession. As we will show, one of the coupled boost vectors will precess around the other's arrow, or *vice versa*. However, while the Wigner-Thomas rotation is interpreted in the configuration space, we apply it to transformations projected in abstract gauge fields.

$$R = \begin{bmatrix} \cos\vartheta + (1-\cos\vartheta)u_1^2 & (1-\cos\vartheta)u_1u_2 & (1-\cos\vartheta)u_1u_3 & iu_1\sin\vartheta \\ (1-\cos\vartheta)u_2u_1 & \cos\vartheta + (1-\cos\vartheta)u_2^2 & (1-\cos\vartheta)u_2u_3 & iu_2\sin\vartheta \\ (1-\cos\vartheta)u_3u_1 & (1-\cos\vartheta)u_3u_2 & \cos\vartheta + (1-\cos\vartheta)u_3^2 & iu_3\sin\vartheta \\ -iu_1\sin\vartheta & -iu_2\sin\vartheta & -iu_3\sin\vartheta & 1 \end{bmatrix} +$$

$$+ \begin{bmatrix} 0 & -u_3\sin\vartheta & u_2\sin\vartheta & -iu_1\sin\vartheta \\ u_3\sin\vartheta & 0 & -u_1\sin\vartheta & -iu_2\sin\vartheta \\ -u_2\sin\vartheta & u_1\sin\vartheta & 0 & -iu_3\sin\vartheta \\ iu_1\sin\vartheta & iu_2\sin\vartheta & iu_3\sin\vartheta & 0 \end{bmatrix} \quad (6)$$

Now, we see that

$$R = \Lambda\cos\vartheta + \begin{bmatrix} 0 & -u_3 & u_2 & -iu_1 \\ u_3 & 0 & -u_1 & -iu_2 \\ -u_2 & u_1 & 0 & -iu_3 \\ iu_1 & iu_2 & iu_3 & 0 \end{bmatrix}\sin\vartheta = \frac{\Lambda}{\kappa} + \frac{1}{c}\begin{bmatrix} 0 & -v_3 & v_2 & -iv_1 \\ v_3 & 0 & -v_1 & -iv_2 \\ -v_2 & v_1 & 0 & -iv_3 \\ iv_1 & iv_2 & iv_3 & 0 \end{bmatrix} \quad (7)$$

or inverted

$$\Lambda = \frac{R}{\cos\vartheta} - \begin{bmatrix} 0 & -u_3 & u_2 & -iu_1 \\ u_3 & 0 & -u_1 & -iu_2 \\ -u_2 & u_1 & 0 & -iu_3 \\ iu_1 & iu_2 & iu_3 & 0 \end{bmatrix}\operatorname{tg}\vartheta \quad (8)$$

The matrices in the second terms in (7) and (8) are also rotation-like. They suggest precession of the axis defined by the vector *v* around the $u_i$ velocity components. Thus, the Lorentz transformation (8) is expressed in terms of a real rotation minus a precession, and both are functions of the relative velocity of the interacting agents. In the instance of low velocity, the transformation *R* turns into the traditionally known Lorentz transformation. In the presence of a velocity-dependent field, the transformation $\Lambda$ must be extended according to the rule expressed in (8). The unitary velocity components in the precession matrix can be interpreted also like spatial projections of the velocity-dependent IFCS vectors from a velocity-dependent *field* (**D**) to the configuration *space*.

We recall that the IFC theory requires invariance under the combination of the Lorentz transformation and the hypersymmetry (HySy) of the IFC transformation. One can check in [2], [3] that the HySy can be represented by the so-called tau ($\tau$) algebra that (in $\tau_3$ representation) led to two transformation matrices: *E* and $\tau_3$. Remember:

$$E = \begin{bmatrix} 1 & 0 & 0 & 0 \\ 1 & 0 & 0 & 0 \\ 1 & 0 & 0 & 0 \\ 0 & 0 & 0 & 1 \end{bmatrix} = \begin{bmatrix} I_L & 0 \\ 0 & 1 \end{bmatrix} \text{ and } \tau_3 = \begin{bmatrix} 1 & 0 & 0 & 0 \\ 1 & 0 & 0 & 0 \\ 1 & 0 & 0 & 0 \\ 0 & 0 & 0 & -1 \end{bmatrix} = \begin{bmatrix} I_L & 0 \\ 0 & -1 \end{bmatrix},$$

where $I_L$ is a [3x3] minor matrix, introduced in [2]. (*E* is the unit element of the HySy group.)

According to the above derivations, in the presence of a velocity-dependent field, one should apply the extended $R$ transformation matrix (cf., Eqs. (4), (7)) for the matrix of the Lorentz transformation. Let's introduce a [3x3] minor matrix in $R$, to get $R$ in the form

$$R = \begin{bmatrix} R^{(M3)} & 0 \\ 0 & 1 \end{bmatrix}$$

where (cf., (6)):

$$R^{(M3)} = \begin{bmatrix} \cos\vartheta+(1-\cos\vartheta)u_1^2 & (1-\cos\vartheta)u_1u_2 - u_3\sin\vartheta & (1-\cos\vartheta)u_1u_3 + u_2\sin\vartheta \\ (1-\cos\vartheta)u_2u_1 + u_3\sin\vartheta & \cos\vartheta+(1-\cos\vartheta)u_2^2 & (1-\cos\vartheta)u_2u_3 - u_1\sin\vartheta \\ (1-\cos\vartheta)u_3u_1 - u_2\sin\vartheta & (1-\cos\vartheta)u_3u_2 + u_1\sin\vartheta & \cos\vartheta+(1-\cos\vartheta)u_3^2 \end{bmatrix} \quad (9)$$

Transformations $T^{(D)}$ in a velocity-dependent (**D**) field under the combination of the extended Lorentz transformation and the transformation matrices of the HySy take the following forms:

$$T_+^{(D)} = E \cdot R = \begin{bmatrix} I_L & 0 \\ 0 & 1 \end{bmatrix} \begin{bmatrix} R^{(M3)} & 0 \\ 0 & 1 \end{bmatrix} = \begin{bmatrix} I_L \cdot R^{(M3)} & 0 \\ 0 & 1 \end{bmatrix} \quad \text{and}$$

$$T_-^{(D)} = \tau_3 \cdot R = \begin{bmatrix} I_L & 0 \\ 0 & -1 \end{bmatrix} \begin{bmatrix} R^{(M3)} & 0 \\ 0 & 1 \end{bmatrix} = \begin{bmatrix} I_L \cdot R^{(M3)} & 0 \\ 0 & -1 \end{bmatrix} \quad (10)$$

Now, we can formulate the sought-after transformation of the field (convolution of a traditional SM field and the associated non-SM **D** field) that is expected to eliminate "unwanted" masses of the quanta ($\delta$) of the **D** field. Note again that in contrast to the fixed mass of all SM bosons, the mass of $\delta$ depends on the relative velocity between the two interacting isotopes of the concerned field-charges. Therefore, we are expecting a transformation formula depending on velocity.

There are two bosons mediating interactions in these fields. There appears one of the SM (plus gravity) bosons, depending on the kind of interaction that we denote by a general character $\xi$. ($\xi$ may denote either the graviton or the photon, one of the weak vector bosons, or one of the strong gluons.) There appear also the bosons of the **D** field, $\delta$.

Thus, the transformation of their fields may take the following two forms in each of the respective interactions:

$$\begin{bmatrix} \mathbf{D}' \\ \mathbf{X}_{SM}' \end{bmatrix} = T_+^{(D)} \begin{bmatrix} \mathbf{D} \\ \mathbf{X}_{SM} \end{bmatrix} = \begin{bmatrix} I_L \cdot R^{(M3)} & 0 \\ 0 & 1 \end{bmatrix} \begin{bmatrix} \mathbf{D} \\ \mathbf{X}_{SM} \end{bmatrix} \quad (11)$$

and

$$\begin{bmatrix} \mathbf{D}' \\ \mathbf{X}_{SM}' \end{bmatrix} = T_-^{(D)} \begin{bmatrix} \mathbf{D} \\ \mathbf{X}_{SM} \end{bmatrix} = \begin{bmatrix} I_L \cdot R^{(M3)} & 0 \\ 0 & -1 \end{bmatrix} \begin{bmatrix} \mathbf{D} \\ \mathbf{X}_{SM} \end{bmatrix} \quad (12)$$

We can see that these transformations do not affect any of the SM bosons ($\mathbf{X}'_{SM}$ coincide with $\mathbf{X}_{SM}$, $\xi'$ are equal to $\xi$). The latter are not subjects to any transformation in the **D** field[5]. It is reassuring that the isotopic field-charge model does not destroy the SM, it only extends it at very high velocities of the interacting agents. Eqs. (11) and (12) rotate the non-SM **D** field of the massive intermediate bosons and one of the $\mathbf{X}_{SM}$ SM fields to produce Goldstone bosons consisting of massless IFC bosons in **D'**, and the respective SM bosons in $\mathbf{X'}_{SM}$ (cf, footnote 2).

---

[5] The Weinberg angle mixes two bosons appearing in the same interaction field. The CKM angles mix quark flavours in another, but also SM field. The HySy rotation angle affects an SM and a non-SM field. It does not mix

The transformation of the **D** section of the field is the same both in (11) and (12)

$$\mathbf{D'} = I_L \cdot R^{(M3)} \cdot \mathbf{D} .$$

Since $I_L \cdot R^{(M3)}$ is a [3x3] matrix, this can be written as

$$\begin{bmatrix} \mathbf{D}_1' \\ \mathbf{D}_2' \\ \mathbf{D}_3' \end{bmatrix} = I_L \cdot R^{(M3)} \begin{bmatrix} \mathbf{D}_1 \\ \mathbf{D}_2 \\ \mathbf{D}_3 \end{bmatrix}$$

Let's write the transformation of this section of the field in detail. First, investigate the $I_L \cdot R^{(M3)}$ product:

$$I_L \cdot R^{(M3)} =$$

$$= \begin{bmatrix} 1 & 0 & 0 \\ 1 & 0 & 0 \\ 1 & 0 & 0 \end{bmatrix} \begin{bmatrix} \cos\vartheta + (1-\cos\vartheta)u_1^2 & (1-\cos\vartheta)u_1 u_2 - u_3 \sin\vartheta & (1-\cos\vartheta)u_1 u_3 + u_2 \sin\vartheta \\ (1-\cos\vartheta)u_2 u_1 + u_3 \sin\vartheta & \cos\vartheta + (1-\cos\vartheta)u_2^2 & (1-\cos\vartheta)u_2 u_3 - u_1 \sin\vartheta \\ (1-\cos\vartheta)u_3 u_1 - u_2 \sin\vartheta & (1-\cos\vartheta)u_3 u_2 + u_1 \sin\vartheta & \cos\vartheta + (1-\cos\vartheta)u_3^2 \end{bmatrix} =$$

$$= \begin{bmatrix} \cos\vartheta + (1-\cos\vartheta)u_1^2 & (1-\cos\vartheta)u_1 u_2 - u_3 \sin\vartheta & (1-\cos\vartheta)u_1 u_3 + u_2 \sin\vartheta \\ \cos\vartheta + (1-\cos\vartheta)u_1^2 & (1-\cos\vartheta)u_1 u_2 - u_3 \sin\vartheta & (1-\cos\vartheta)u_1 u_3 + u_2 \sin\vartheta \\ \cos\vartheta + (1-\cos\vartheta)u_1^2 & (1-\cos\vartheta)u_1 u_2 - u_3 \sin\vartheta & (1-\cos\vartheta)u_1 u_3 + u_2 \sin\vartheta \end{bmatrix}$$

(13)

One sees that the three rows of the resulting matrix coincide ($\mathbf{D}_1' = \mathbf{D}_2' = \mathbf{D}_3' = \mathbf{D'}$), as expected. Thus:

$$\mathbf{D'} = \left[ \cos\vartheta + (1-\cos\vartheta)u_1^2 + (1-\cos\vartheta)u_1 u_2 - u_3 \sin\vartheta + (1-\cos\vartheta)u_1 u_3 + u_2 \sin\vartheta \right] \mathbf{D} \text{ or}$$
$$\mathbf{D'} = \left[ \cos\vartheta + (1-\cos\vartheta)u_1(u_1 + u_2 + u_3) + \sin\vartheta(u_2 - u_3) \right] \mathbf{D}$$

(14)

To discuss this value, let's introduce polar co-ordinates for the $u_i$ unitary projected velocity components of the IFCS. Note that we still have prescribed no constraint for the axes in the configuration space, where the $u_i$ velocity component projections point. We were free to orient those axes arbitrary. Now, let us define the axes so that the inclination angle of **v** in respect of $u_3$ be $\Theta_D$, and let $\psi$ denote the rotation angle around $u_3$ in the $u_1$-$u_2$ plane that is perpendicular to $u_3$.

---

the $\delta$ bosons of the **D** field with any of the SM bosons (denoted by $\xi$), what latter appear simultaneously in one of the SM interaction fields ($\mathbf{X}_{SM}$). Therefore, HySy *does not mix* them, instead it is expected to *rotate* the $\delta$ boson's field while leaving the respective $\xi$ SM boson's field unchanged. The rotation formula to be derived in the following part of the paper corresponds to this expectation.

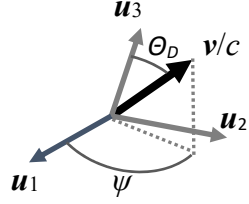

Figure 1

Since $u_i$ are components of a unitary vector, the unitary projections of $v$ are

$u_1 = \sin\Theta_D \cos\psi$
$u_2 = \sin\Theta_D \sin\psi$
$u_3 = \cos\Theta_D$

As can be read from Fig.1, $\Theta_D$ is the angle of the precession of $v$ in a fixed reference frame. $\Theta_D$ characterises the velocity dependence of the transformation of the field **D**. Now:

$$\mathbf{D'} = \begin{bmatrix} \cos\vartheta + (1-\cos\vartheta)(\sin^2\Theta_D \cos^2\psi + \sin^2\Theta_D \cos\psi \sin\psi + \sin\Theta_D \cos\psi \cos\Theta_D) + \\ +\sin\vartheta(\sin\Theta_D \sin\psi - \cos\Theta_D) \end{bmatrix} \mathbf{D}$$

One can fix the reference frame, considering that the precession of $v$ around the axis $u_3$ cannot depend on the phase angle (of a rotation by $\psi$) in the $u_1$-$u_2$ plane. One can interpret $\psi$ as a phase parameter of the spontaneous symmetry breaking in the **D** field. So, we are free to fix $\psi$ (by an arbitrary choice) as $\psi = \pi/2$. In this case $\cos\psi = 0$, $\sin\psi = 1$. With the above assumptions on the orientation of the reference frame of the velocity components:

$$\mathbf{D'} = [\cos\vartheta + \sin\vartheta(\sin\Theta_D - \cos\Theta_D)]\mathbf{D}$$

Considering the identity $\cos\Theta_D - \sin\Theta_D \equiv \sqrt{2}\cos(\Theta_D + \frac{\pi}{4})$:

$$\mathbf{D'} = \left[\cos\vartheta - \sin\vartheta\sqrt{2}\cos(\Theta_D + \frac{\pi}{4})\right]\mathbf{D}$$

The transformation of **D** is a function of the symbolic angles $\vartheta$ and $\Theta_D$. Both can be expressed with the relative velocity of the interacting field-charges. $\vartheta$ is defined by the Lorentz transformation, $\Theta_D$ by the transformation in the HySy field **D**. In simpler form:

$$\mathbf{D'} = \left[\sqrt{1-\left(\frac{v}{c}\right)^2} - \sqrt{2}\frac{v}{c}\cos(\Theta_D + \frac{\pi}{4})\right]\mathbf{D} \qquad (15)$$

Inserting this in (11) [and (12), respectively]:

$$\begin{bmatrix} \mathbf{D'} \\ \mathbf{X}_{SM}' \end{bmatrix} = \begin{bmatrix} \sqrt{1-\left(\frac{v}{c}\right)^2} - \sqrt{2}\frac{v}{c}\cos(\Theta_D + \frac{\pi}{4}) & 0 \\ 0 & \pm 1 \end{bmatrix} \begin{bmatrix} \mathbf{D} \\ \mathbf{X}_{SM} \end{bmatrix} \qquad (16)$$

According to (16), $\vartheta$ and $\Theta_D$ define together the transformation that eliminates unwanted masses produced by the spontaneous symmetry breakdown in the **D** field and justify the mass of the $\delta$ boson. This formula complies with the transformation of the electro-weak field by the Weinberg mixing angle. However, there are differences as well.

Firstly, (16) transforms two coupled fields together, one of which is not a SM field.

Secondly, while there are fixed mass bosons in the weak interaction, the mass of the HySy field's boson depends on the relative velocity of the interacting field-charges. This is in keeping with the velocity dependence of the **D** field and is reflected in the field's transformation formula for the elimination of unwanted masses produced by the spontaneous symmetry breakdown.

Thirdly, we must remark that the assumption of isotopic field-charges originates in the asymmetry expressed in the Møller scattering matrix [25], [7]. That assumption involved the mass difference between the IFC siblings. A later formula obtained in the SM for the Møller scattering asymmetry for electrons includes the (weak mixing) Weinberg angle. (The weak mixing angle explains only the surplus of the $Z^0$ boson. The BEH mechanism gives account of the full mass of $Z^0$.) The value of the Weinberg angle varies depending on the momentum transfer. The momenta affect the fixed masses of the related weak bosons. In contrast to that, although the transformation formula in the **D** field affects also the mass of the quanta of the field, it leads to a boson mass depending on the relative velocity of the particles between which it mediates. Moreover, the appearance of the velocity-dependent angle in the formula for the transformation of the **D** field is simpler than in the Møller scattering asymmetry. At the same time, the angle in (16) runs over a wider scale than the Weinberg angle does.

In short, Eq. (16) is the formula by which spontaneous symmetry breaking transforms the respective quanta of the original SM field and the **D** field.

We can expect the (11)-(12) rotation matrices in the **D**-**X**$_{SM}$ field couple (as expressed in (16)) in a rotation matrix form

$$\begin{bmatrix} \sqrt{1-\left(\frac{v}{c}\right)^2} - \sqrt{2}\frac{v}{c}\cos(\Theta_D + \frac{\pi}{4}) & 0 \\ 0 & \pm 1 \end{bmatrix} = \begin{bmatrix} \cos\varphi(v,\Theta_D) & \sin\varphi(v,\Theta_D) \\ -\sin\varphi(v,\Theta_D) & \cos\varphi(v,\Theta_D) \end{bmatrix} \quad (17)$$

where $\varphi(v,\Theta_D)$ denotes an angle that mixes the **D** and the respective **X**$_{SM}$ fields. However, the inclination angle $\Theta_D$ appears to be more characteristic for the rotation of the **D** field than $\varphi$.

## 4 Discussion of the resulted field transformation

According to (17) $\sin\varphi(v,\Theta_D)=0$, involves meaning stable values for $\varphi(v,\Theta_D)$, while $v$ and $\Theta_D$ may vary. Since $\sin\varphi(v,\Theta_D)=0$,

(a) $\varphi = 0$ and $\cos\varphi = 1$; or
(b) $\varphi = \pi$ and $\cos\varphi = -1$.

In case (a): $\varphi = 0$ there occurs no transformation in the field **D**. In this case (cf., (11)) $T_+^{(D)}$ turns into the identity transformation: $T_+^{(D)} = \begin{bmatrix} 1 & 0 \\ 0 & 1 \end{bmatrix}$.

The case (b): $\varphi = \pi$ corresponds to a real transformation, by the matrix $\tau_3$. According to (12) and (17)

$$\cos(\Theta_D + \frac{\pi}{4}) = \frac{\sqrt{1-(v/c)^2}+1}{\sqrt{2}(v/c)}. \tag{18}$$

The formula in (18) provides limits for the domain of interpretation of $v/c$, and respectively, for $\Theta_D$. One must avoid having the right side becoming larger than the value allowed by a cosine function. Discussion of these limits allows us to define the limits where HySy is broken. At first, we exclude $\Theta_D$ outside the domain $-\frac{\pi}{2} \leq \theta_D \leq \frac{\pi}{2}$ (otherwise the projection of $v$ would point in the opposite direction than $v_3$), and we also consider that $v/c$ runs from 0 to 1.

The negative value of the square root in the numerator in (18) provides either – excluded – precession angles less than $-\pi/2$ for $\Theta_D$, or positive precession angles. Furthermore, for $\varphi = \pi$ rotation of the **D**-**X**$_{SM}$ fields causes a sign inversion in the $v$-$\Theta_D$ plane [flips the (velocity) vectors in these fields over in the opposite direction], only negative $\Theta_D$ precession angles can be interpreted: $-\frac{\pi}{2} \leq \theta_D \leq 0$. Thus, all precession angles provided by negative square roots in the numerator should be excluded.

Considering the positive values of the numerator, expression (18) is meaningless when $\frac{v}{c} < \sqrt{\frac{8}{9}} \cong 0{,}943$. This is the minimum velocity where HySy prevails. Below this limit velocity, $v < \sqrt{8/9}\,c$, the HySy is broken. At the critical $v = \sqrt{8/9}\,c$, the precession angle $\Theta_D = -\frac{\pi}{4}$. Starting from this value of the spontaneous symmetry breaking angle value, while the velocity (kinetic energy) increases further, the value of the $\Theta_D$ precession angle spontaneously bifurcates. It either increases, reaching $\Theta_D = 0$ at $v = c$; or decreases, reaching $\Theta_D = -\frac{\pi}{2}$ at $v = c$. The observable domain of $\Theta_D$ varies between $-\frac{\pi}{2} < \theta_D < 0$ (cf., Figs. 2 and 3).

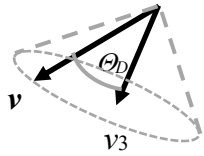

Figure 2

In other words, according to (17) the HySy rotation angle is $\varphi = \varphi(v, \Theta_D)$, as learned in the SM. The identity transformation (case (a)) indicates no transformation of the field **D** in SM terms. This case says field **D** is present at the range $0 < v \leq c$, but its presence does not guarantee that HySy phenomena, like a HySy boson (dion), can be observed.

There may show up domains where HySy is broken. The real transformation (case (b)) indicates a $\pi$ angle rotation of the plane of the fields **D-X**$_{SM}$, i.e., vectors flip over in opposite direction. This justifies negative precession angles around velocity vectors in the **D** field.

The fixed value of the $\varphi$ rotation angle in the SM hides the essence of the rotation of a field beyond the SM[6], like **D**. Since **D** exists beyond the SM, new rules may prevail in it. The essential characteristic angle is hidden in the velocity dependence of $\varphi(v,\Theta_D)$. The spontaneous symmetry breaking in HySy is characterised by that $\Theta_D$ precession angle. The curve of *v/c* in the function of the available values for $\Theta_D$ shows a sombrero-like graph (cf. Fig. 3). This complies with similar shapes for SM spontaneous symmetry breakings. According to the discussion of the set of values for the Eq. (18), $\Theta_D$ is interpreted between $-\pi/2$ and $0$ (indicated as a bold line), and the respective values of velocity between $\sqrt{8/9} \leq \frac{v}{c} < 1$, at least, considering the positive value of the square root in the numerator in (18). The $\sqrt{8/9} \leq \frac{v}{c}$ limit for *v/c* means the limit under which velocity the HySy is broken.

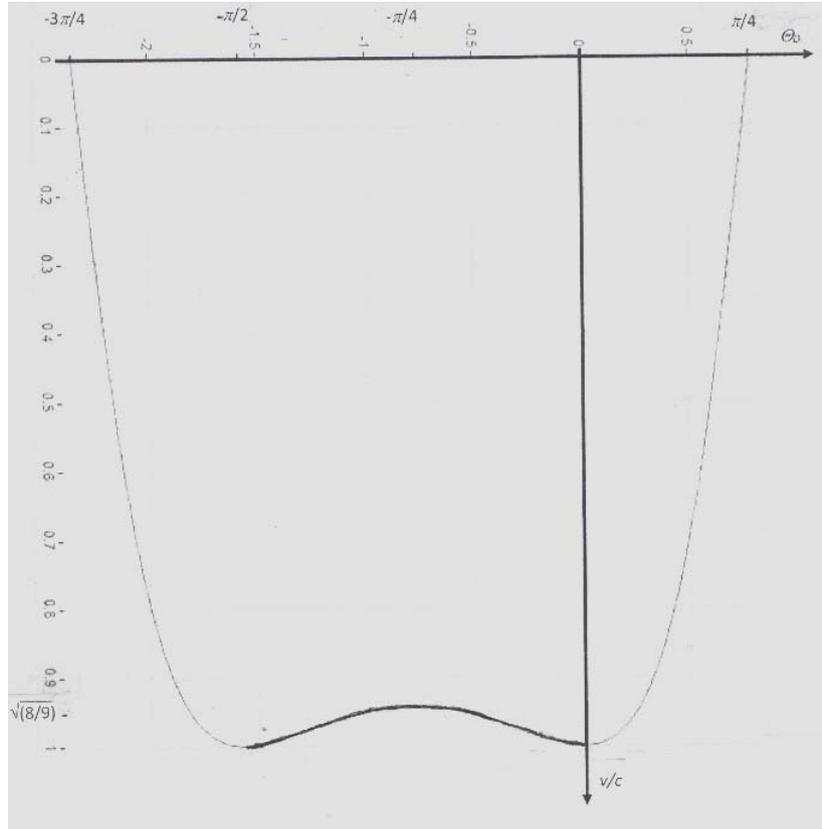

Figure 3: Real velocity.

In point of fact, HySy is effective between velocities $\sqrt{8/9} \leq \frac{v}{c} < 1$ with a domain of the precession angle $-\frac{\pi}{2} < \Theta_D < 0$. (We exclude the boundaries of the domain because no massive particle can appear at velocity *c*.)

---

[6] [24] discusses „the first clear evidence for physics beyond the Standard Model".

The trigonometric formula demonstrates spectacularly the *precession* of the **D** field depending on the velocity (cf., (15)). The angle of the precession expresses the relation of *v* to $v_3$. It can be shown clearly that the *v* vector of the velocity of one of the interacting isotopic field-charges - in respect to the other field-charge - precesses around the straight, marked out by $v_3$. The direction of $v_3$ points between the two, interacting field-charges and does not coincide with the direction of the real velocity *v* (cf., Fig. 1). In another view, the *v* velocity vector precesses around the projection of the third component of the unitary length IFCS from the **D** *field* to the configuration *space*. The angle of this precession changes with the change of the respective velocity *v*. It can be also shown that the vector of velocity is always tangential to the line connecting the interacting agents, but this line is curved in the gauge field **D** induced by their own velocity. The latter can be expressed by the inverse of the matrix in Eq. (16).

The vector *v* is interpreted in the velocity-dependent gauge field **D**. In the chosen orientation of the reference frame in **D**, the direction of its projection to the third axis, marked by $u_3$ (cf., Fig. 1), coincides with the direction of the isotopic field-charge spin. Fig. 2 illustrates an angle-preserving projection of *v* and its third component $v_3$ in the configuration space. The configuration space vector *v* (delineated in Fig. 3), which is tangential to the trajectory of the moving particle, is its *real velocity*. At the same time, the vector component $v_3$, inclined by an angle $\Theta_D$ in respect to *v*, defines the *effective velocity* of a boson-emitting particle in the direction towards a target particle that it enters in interaction with. This effective value is $v_3 = v \cdot \cos \Theta_D$ (cf. Fig. 2). $v_3$ is the longitudinal component of *v*. The transversal component of *v* ($v \cdot \sin \Theta_D$) arrows in an arbitrary direction in the $u_1$-$u_2$ plane, according to a spontaneous precession. Fig. 4 shows the value of $v_3$ as a function of the inclination angle $\Theta_D$. According to Fig. 3 (as a result of Eq. (18)) the HySy enters the scene at the velocity appearing by the angle $\Theta_D = -\pi/4$. As it can be read from Fig. 4, the effective velocity $v_3$ becomes $(2/3)c$ at that point. This means one can observe HySy above the *effective velocity* $v_3 = (2/3)c$.

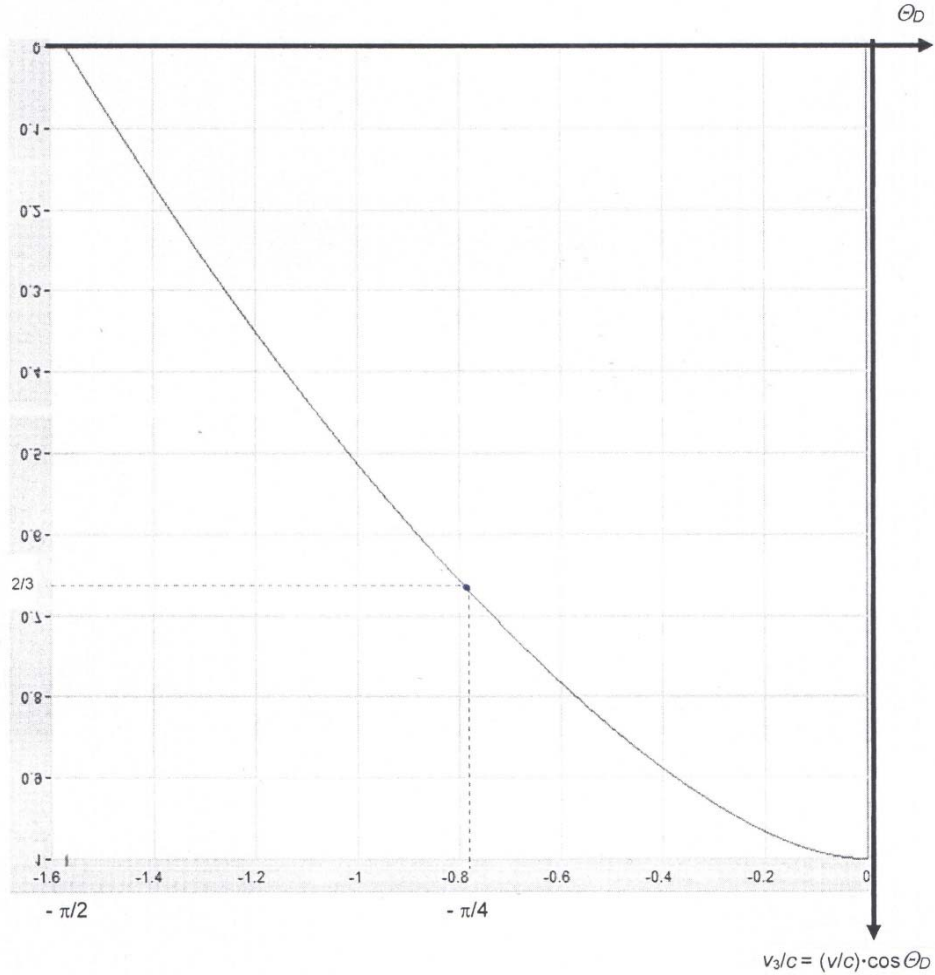

Figure 4: Effective velocity.

## 5 The mass of the mediating boson δ in light of the transformation of the D field

The set of the $\Theta_D$ angle values obtained characterises the rotation of the **D** field that eliminates unwanted masses produced by the spontaneous symmetry breaking, that is responsible for the mass of the field's $\delta$ boson. It is easy to see that the velocity dependence of $\Theta_D$ (18) is in close relation with the velocity-dependent coefficient ($\kappa$) of the mass of the $\delta$ boson.

As we saw, $m(\delta) = m_T - m_V = (\kappa - 1) m_V$ where the value of $m_V$ is equal to the rest mass. At the minimum energy of the appearance of the HySy $\dfrac{m(\delta)}{m_V} = \kappa - 1 = \dfrac{1 - \sqrt{1 - (v/c)^2}}{\sqrt{1 - (v/c)^2}} = 2$. This energy value corresponds to the middle apex in the sombrero curve at $-\pi/4$ (cf., Fig. 3). Accordingly, the $m_T$ mass of the Lorentz boosted isotopic field-charge at the lower limit of HySy should be $\dfrac{m_T}{m_V} = \kappa = 3$. In other words, HySy is broken until the mass of the respective mediating boson does not reach the double of the rest mass of the emitting particle, or, what is the same, the Lorentz boosted mass does not reach the triple of the rest mass of the emitting particle. This expresses the lower limit of the *observability* of a boson $\delta$ that appears at velocity $v = \sqrt{8/9}\,c$ and above.

When $v$ approaches zero, the mass of $\delta$ approaches to 0. However, $\delta$ cannot be observed near such a low velocity, due to the spontaneous symmetry breaking of the IFCS field. This is expected since low velocity means to return to the full domination of the SM. There is not allowed to appear a measurable value for the **D** field's strength and transformation of a boson in the **D** field within the limits of the SM. This means the calculation confirms that one can observe no $\delta$ bosons when a **D** field vanishes. (Moreover, we showed in the previous paragraphs that there exists a stronger exact limit for $v$ in order to eliminate the observation of a massive $\delta$.) This fact confirms that the IFC theory extends the SM so that the SM is left intact and holds at the range of its validity, i.e., at not extremely high energies.

We were seeking to find a transformation of the **D** field that may eliminate the mass of the $\delta$ boson.[7] This is equivalent to a rotation of the field demanded by the spontaneous symmetry breaking and precession of the velocity $v$ around its third projection in the **D** field (that produced the mass of the field's bosons).

## 6 Conclusions

We have derived the transformation formula that eliminates "unwanted" masses produced by the spontaneous symmetry breaking in the **D** field and justified the mass of the quantum of the **D** field. The derivation justifies that **D** must be a gauge field, i.e., velocity dependence cannot be considered a simple rotation in the configuration space defined in the matter field. Earlier publications by the author [3] showed that this **D** field is subject to an invariance under rotations of an $SU(2)$ isomorphic group that characterises hypersymmetry. We demonstrated that the (isotopic field-charge spin) transformation in the **D** field must be coupled with a SM interaction field, and also that the transformation leaves the mediating bosons of the respective SM field intact [10]. The derived formula confirms that the **D** field causes no observable effects at low velocities, but it should be taken into account at relativistic high velocities: it extends the SM but does not influence it in the range of its validity.

We derived a limit velocity $v/c = 2\sqrt{2}/3$, below which HySy is definitely broken. The Lorentz invariance is extended over this limit velocity (energy) by invariance under HySy. A non-SM transformation of the **D** field interpreted by the BEH mechanism and discussed in section 4 justifies the mass of the quanta of the field. This transformation is characterized by a mixing angle $\pi$ and a precession inclination angle $\Theta_D(v)$. This inclination angle of the precession of vectors, interpreted in the velocity-dependent field, rotates the field that is responsible for eliminating the mass of the field's intermediate boson. The latter angle is interpreted by HySy, beyond the SM. The mass of the quanta of the **D** field ($\delta$) depends on $\Theta_D(v)$.

---

[7] Let's avoid confusing the identity transformation of the boson $\delta$ at zero velocity with the field rotation transforming its mass into 0.